\documentclass[	DIV=calc,%
							paper=a4,%
							fontsize=11pt,
							twocolumn,
                            ]{scrartcl}	 					

\usepackage[english]{babel}										
\usepackage[protrusion=true,expansion=true]{microtype}				
\usepackage{enumitem}

\usepackage{amsmath,amsfonts,amsthm}					
\usepackage[pdftex]{graphicx}									
\usepackage[svgnames]{xcolor}									

\usepackage{fancyhdr}												
\usepackage{lastpage}	

\usepackage{enotez}
\setenotez{list-style=itemize, backref=true}
\let\footnote=\endnote

\usepackage{caption}
\fancypagestyle{plain}{} 
\definecolor{myblue}{HTML}{12274d}
\definecolor{mygold}{HTML}{a85c04}
\definecolor{mygrey}{HTML}{A3A9AA}
\definecolor{Indigo}{HTML}{500472} 
\definecolor{Bole}{HTML}{784F41} 
\definecolor{Tiffany}{HTML}{79cbb8}
\definecolor{Isabelline}{HTML}{F1E8E4}

\definecolor{Chocolate}{HTML}{420C14} 
\definecolor{PolyGreen}{HTML}{214e34}
\definecolor{Tiger}{HTML}{A85C04} 
\definecolor{Ivory}{HTML}{EEF3E8}
\definecolor{Puce}{HTML}{C08497}
\definecolor{YInMn}{HTML}{26547C}
\definecolor{Gamboge}{HTML}{F06543}
\colorlet{mygray}{mygrey!75!black}
\usepackage{afterpage}

\colorlet{titlecolor}{myblue}
\colorlet{linecolor}{mygold}
\colorlet{affcolor}{mygray}
\colorlet{titlecolor}{Black}
\colorlet{linecolor}{Grey}
\colorlet{affcolor}{Grey}

\usepackage{charter} 
\def\fonttitle{LinuxBiolinumT-OsF}

\def\fontauthors{LinuxBiolinumT-OsF}

\newcommand{\formatauthor}[1]{
\fontfamily{\fontauthors}\selectfont
\color{titlecolor}#1}
\newcommand{\formataffil}[1]{&\footnotesize \fontfamily{\fontauthors}\selectfont\color{affcolor}#1\\[-6pt]}
\usepackage{titling}															

\newcommand{\HorRule}{\color{linecolor}
									  	\rule{\linewidth}{1pt}%
										}
\pretitle{\vspace{-30pt} \begin{center} \HorRule\\ 
				\fontsize{26}{30}\fontfamily{\fonttitle}\selectfont
    \bfseries
    \color{titlecolor} 
				}
\title{What if automating AI R\&D triggers an intelligence explosion?}					
\posttitle{\par\end{center}\vskip -0.5em}

\preauthor{\hspace{2.0cm}\begin{tabular}{p{3.5cm}p{0.65\linewidth}}
}

\author{
\formatauthor{Alan Chan*}
\formataffil{GovAI}
\formatauthor{Christoph Winter}
\formataffil{CASP, University of Cambridge, Institute for Law \& AI}
\formatauthor{Andrew Barto}
\formataffil{University of Massachusetts Amherst}
\formatauthor{Jakub Pachocki}
\formataffil{OpenAI}
\formatauthor{Geoffrey Hinton}
\formataffil{University of Toronto, Vector Institute}
\formatauthor{Eric Horvitz}
\formataffil{Microsoft}
\formatauthor{Yoshua Bengio}
\formataffil{Mila (Quebec AI Institute), Université de Montréal, LawZero}
\formatauthor{Dawn Song}
\formataffil{University of California, Berkeley}
\formatauthor{Jack Clark}
\formataffil{Anthropic}
\formatauthor{Hilary Greaves}
\formataffil{University of Oxford}
\formatauthor{Anton Korinek}
\formataffil{University of Virginia, Anthropic}
\formatauthor{Samuel Hammond}
\formataffil{Foundation for American Innovation}
\formatauthor{Thore Graepel}
\formataffil{University College London}
\formatauthor{Ben Bariach}
\formataffil{University of Oxford}
\formatauthor{Philip H. S. Torr}
\formataffil{University of Oxford}
\formatauthor{Sheila A. McIlraith}
\formataffil{University of Toronto, Vector Institute}
\formatauthor{Jeff Clune}
\formataffil{University of British Columbia, Vector Institute}
\formatauthor{Sam Manning}
\formataffil{GovAI, Foundation for American Innovation}
\formatauthor{Girish Sastry}
\formataffil{Guidelight}
\formatauthor{Tom Davidson}
\formataffil{Forethought}
\formatauthor{Daniel Eth}
\formataffil{AI Policy Institute}
\formatauthor{Sören Mindermann*}
\formataffil{CASP, University of Cambridge}
}											
\postauthor{ 								
					\end{tabular}\par\vspace{3mm}\HorRule\vspace*{-12mm}}     
\date{}																				

\usepackage[backend=biber, bibstyle=ieee, citestyle=numeric-comp,
  natbib=true, sorting=none, labeldateparts, backref=true,
  maxbibnames=99, url=false, isbn=false, maxcitenames=2, mincitenames=1]{biblatex}  
\PassOptionsToPackage{hyphens}{url}
\AtEveryBibitem{\clearlist{language}}
\renewbibmacro*{date}{%
  \iffieldundef{year}
    {\bibstring{nodate}}
    {\printdate}}

\usepackage{hyperref}
\hypersetup{
colorlinks   = true, 
linkcolor=black,
citecolor=YInMn,
urlcolor=mygrey!75!black,
}
\usepackage{xurl}
\usepackage{stfloats}

\begin{document}
\makeatletter
\twocolumn[
   \begin{@twocolumnfalse}
     \maketitle
     \begin{abstract}
     \vspace{-4mm}
     \subsection*{\hspace{0.45\linewidth}{Abstract}}

     \noindent In contrast to even a year ago, AI systems now write most of the code inside the companies that build them. As more of the AI research and development (R\&D) pipeline is automated, could AI progress radically accelerate in an ``intelligence explosion,'' where years of advances are compressed into months or less? Preliminary evidence suggests that it could. In this work, we assess this evidence, analyze an intelligence explosion's potential impacts, and propose policy responses. AI systems are on track to automate most AI R\&D work within a few years, and possibly all of it. If this triggers an intelligence explosion, it could dramatically bring forward AI's benefits, but also pose extreme risks: capabilities growth could accelerate far beyond what society can keep up with, humanity could lose control over superhuman AI systems, and checks on power within and between states, companies, and branches of government could be severely eroded. Although there remains much uncertainty about these possibilities, the high stakes warrant serious further attention. Policymakers should urgently obtain more visibility into the automation of AI R\&D, develop ways to steer and constrain an intelligence explosion, and prepare society to adapt to an intelligence explosion's impacts.
     \end{abstract}
     \vspace{5mm}
    \end{@twocolumnfalse}
]

\makeatother
\thispagestyle{fancy} 			

\section*{Introduction}

\begin{figure*}[!b]
	\centering
	\includegraphics[width=\textwidth]{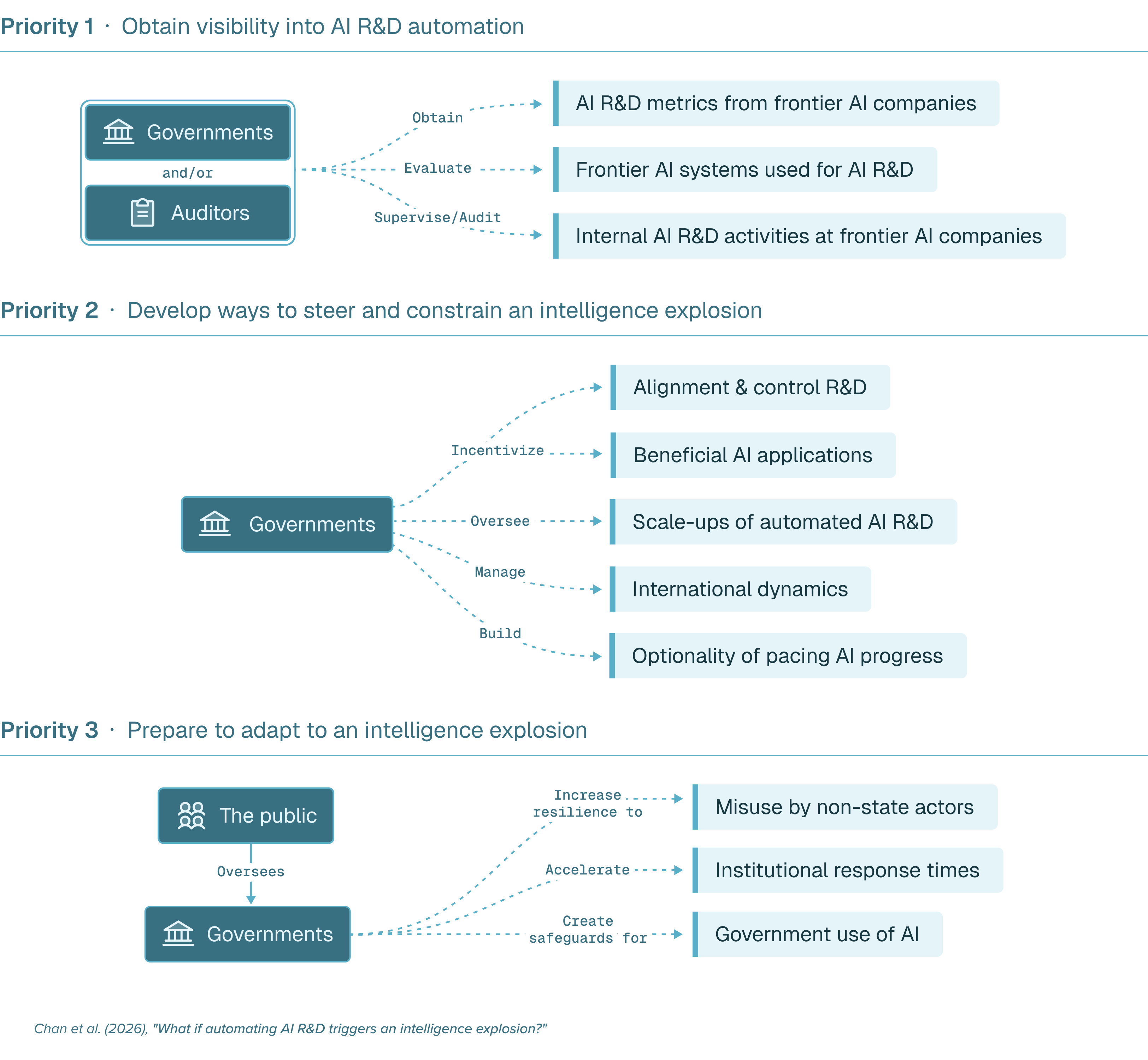}
	\caption{Policymakers should urgently: (1) obtain visibility into automation of AI R\&D within frontier AI companies; (2) develop ways to steer and constrain an intelligence explosion; and (3) prepare to adapt to an intelligence explosion's impacts.}
	\label{fig:priorities}
\end{figure*}

Computer scientists have long theorized that AI systems would one day design ever-better successors, producing systems that rapidly outnumber, outpace, and outperform humans~\citep{good1966speculations, turing_intelligent_1951, evans_agentic_2026}. Today, frontier AI companies are aiming to automate AI R\&D~\citep{pachocki_alien_2026, favaro_when_2026} while deploying more capital as a fraction of U.S. GDP than the Manhattan Project and Apollo Program combined~\citep{tunguz_are_2025}. What happens if they succeed?

We are centrally concerned with the possibility of an \textbf{intelligence explosion: a dramatic AI-driven acceleration of AI progress, compressing advances that would otherwise take years into months or less}. This acceleration would be a qualitative shift from the rapid but largely steady progress of the last few years.

An intelligence explosion could arise from AI contributing to advances in hardware and/or software. Hardware advances increase the quality and quantity of computing hardware (compute) used for developing and running AI systems. Software advances improve the data, algorithms, code, and processes used in AI R\&D; they include more efficient training and inference on existing hardware~\citep{leiserson_theres_2020,ho_algorithmic_2024,openai_how_2026}, improved research management and operations, better synthetic data and training environments~\citep{abdin_phi4_2024,denain_faq_2026}, and novel paradigms in AI~\citep{guo_deepseek_2025,openai_learning_2024}. Software advances warrant particular attention in the near term for two reasons. First, AI systems appear to be rapidly improving at AI R\&D, making them better at producing such advances. Second, software advances allow fast feedback loops: improved AI systems can be redeployed into the R\&D process almost immediately, whereas hardware improvements typically depend on years-long manufacturing and construction cycles. This piece therefore focuses on the possibility of a \textbf{software-driven intelligence explosion}, where automation of AI R\&D drives an intelligence explosion through software advances alone~\citep{eth_will_2025}.

\textbf{Preliminary evidence suggests that a software-driven intelligence explosion is possible.} If one does happen, it could be the most consequential technological development in history: AI systems could rapidly eclipse human experts across most domains and radically accelerate technological progress. Given the stakes and the potentially narrow window for action, we argue that preparing for an intelligence explosion should be an urgent priority, including at the highest levels of government leadership.

\section*{AI is rapidly automating AI R\&D}

AI systems now either assist with or autonomously carry out major parts of the AI R\&D pipeline. In contrast to even just a year ago, R\&D staff at leading AI companies delegate core R\&D tasks to teams of AI systems, and some delegate all coding. Anthropic reports that AI systems' share of approved code rose from low single digits to over 80\% between January 2025 and May 2026~\citep{favaro_when_2026}, while the proportion of R\&D work autonomously completed with only high-level human supervision rose from 1\% to 26\% between March and August 2026~\citep{favaro_measurements_2026}. OpenAI reports that ``AI assistance is used in practically all parts of the company across technical and non-technical teams with code-executing agents used in training, evaluating, and securing future models,'' and Google reports that ``AI is used in almost all work that involves writing code or configuration, technical design, research ideation, to different degrees depending on the task''~\citep{metr_frontier_2026}.

The best AI systems now complete AI R\&D tasks that take human experts hours to days, compared to only being able to complete seconds-long tasks in 2023~\citep{metr_time_2026, openai_research_2026,posttrainbench_2026}. AI systems also sometimes beat human experts: they have autonomously produced better solutions to an AI safety research problem~\citep{wen_automated_2026}, and in some situations predict more accurately which research ideas will pan out~\citep{wen_predicting_2025} and which next steps are worth taking~\citep{favaro_when_2026}. In an early proof of concept, an automated AI research pipeline generated research ideas, ran experiments, and wrote a paper that passed peer review at a workshop held at a top-tier machine-learning venue~\citep{lu_towards_2026}.\footnote{Machine-learning venues typically have a main conference track and several workshop tracks. One caveat to the results is that these workshop tracks can have somewhat laxer standards than the main conference track.}

Today's AI systems still have many weaknesses. They sometimes disobey instructions, cheat on tasks, misrepresent their work, and are unable to complete some tasks at all, necessitating human intervention~\citep{rabanser_towards_2026, anthropic_claude_2026, openai_gpt6_2026}. For example, GPT-6 fails some of OpenAI's research debugging tasks that experienced human researchers can complete (albeit in hours or days)~\citep{openai_gpt6_2026}. Success on benchmarks can also fail to translate into real-world productivity boosts~\citep{whitfill_many_2026}.

Still, AI systems are rapidly improving at AI R\&D. AI R\&D may well be automated before most other work: it is a primarily digital domain with many clear measures of success, automating it would offer a major competitive edge in the AI industry, and AI companies have unmatched data on---and expertise in---their own workflows. Some tentative extrapolations of recent trends suggest that months-long AI R\&D projects will be automated by mid-2028.\footnote{According to the METR time-horizon metric~\citep{metr_measuring_2025, metr_time_2026}, the length of tasks that AI systems can complete initially doubled roughly every 7 months, accelerating to about every 3 months since 2024. Extrapolating this more recent trend would suggest that by mid-2028, AI systems will be able to complete tasks requiring several months of human expert time, well within the range of many AI R\&D projects. See also \citet{kokotajlo_ai_2025}.} Overall, we should expect much more substantial automation of AI R\&D over the next few years, and even full automation within this timeframe should be taken seriously.

\section*{Automating AI R\&D could trigger an intelligence explosion}

The mechanism for a software-driven intelligence explosion has two parts: (1) AI systems expand the effective R\&D workforce as they get better and faster at AI R\&D, and (2) this workforce produces still better AI systems that expand the workforce even further in a recursive feedback loop. Although the existing evidence is preliminary and sometimes mixed, it suggests that this mechanism could radically accelerate AI progress, overcoming frictions such as diminishing returns and hard-to-automate tasks. 

\begin{figure*}[!t]
	\centering
	\includegraphics[width=\textwidth]{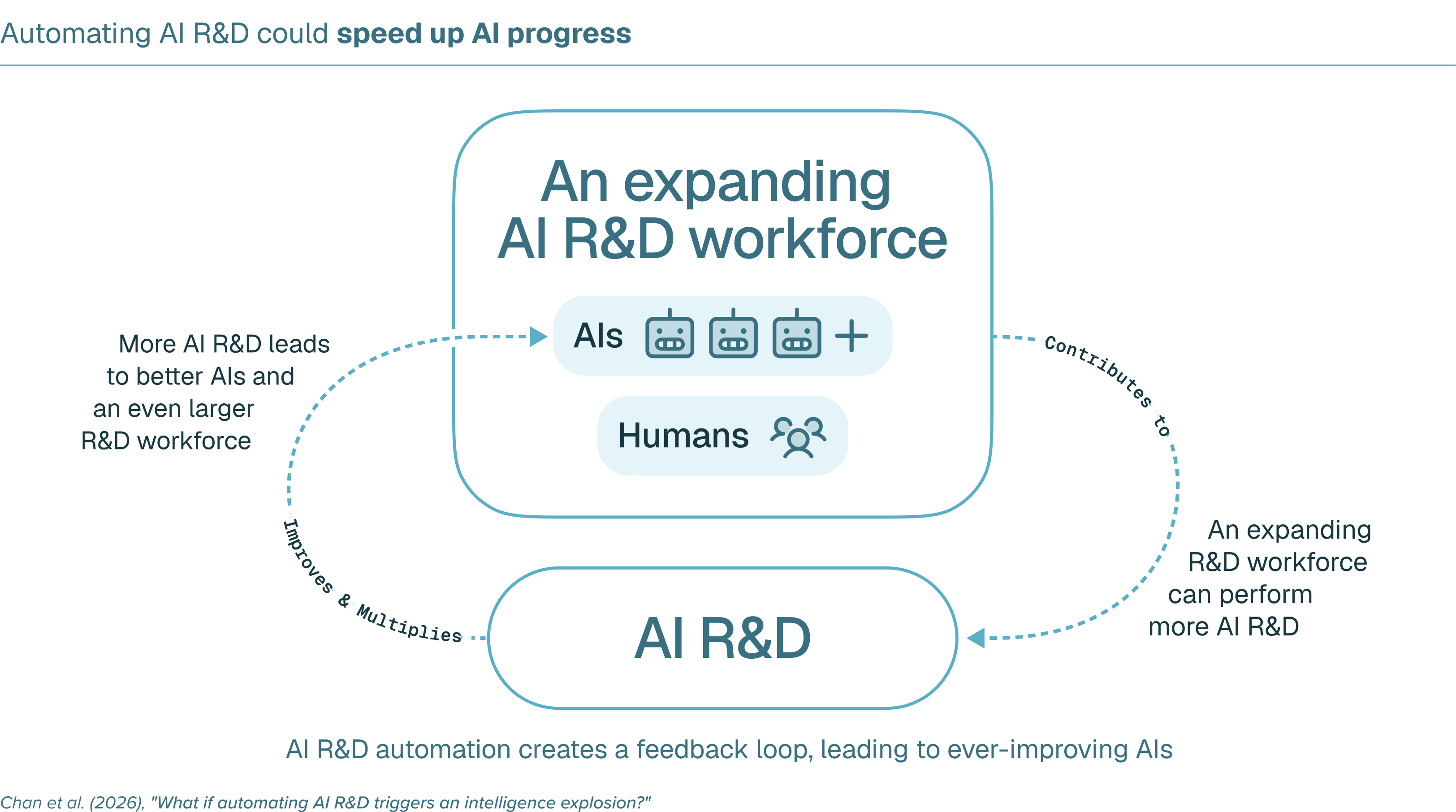}
	\caption{The mechanism for a software-driven intelligence explosion has two parts: (1) AI systems expand the effective R\&D workforce as they get better and faster at AI R\&D, and (2) this workforce produces still better AI systems that expand the workforce even further in a recursive feedback loop.}
	\label{fig:mechanism}
\end{figure*}

\subsection*{The mechanism and frictions}

Each new generation of AI systems will perform a growing range of R\&D tasks faster and better than humans can, effectively yielding a larger, smarter, and faster automated R\&D workforce. Once AI systems reach expert-level AI R\&D capabilities at runtime costs comparable to those of today's systems, the compute available to a single frontier developer today could sustain an AI workforce equivalent to at least millions of top human researchers (see the Supplementary Materials [SM]), dwarfing the thousands of researchers that frontier companies currently employ. The size and duration of any resulting speed-ups remain uncertain and merit further study. Still, to see why they could be substantial, consider the reverse: AI progress would likely slow dramatically if today's human researchers were ten times fewer or slower.

AI systems also help build more capable and efficient successors, further expanding the automated R\&D workforce and creating a feedback loop that could sustain and compound successive speed-ups. Past technologies have also involved feedback loops: for example, better computer chips power better chip design tools. What may distinguish a software-driven intelligence explosion is how much AI systems would contribute to producing the next generation: as they substitute for humans on a growing share of R\&D tasks, each software advance speeds up an ever-larger share of the R\&D pipeline. At full automation, even the current pace of efficiency improvements would grow the automated R\&D workforce 100-fold over months or years,\footnote{\citet{cottier_llm_2025} find that the price of running LLMs to achieve a given capability milestone (e.g., GPT-4-level performance on a math benchmark) has decreased by roughly 9- to 900-fold per year, depending on the capability milestone. While a portion of this cost decline has come from hardware improvements (reducing the cost per computation), a substantial fraction is due to software improvements.} a relative expansion that took the U.S. researcher population seven decades~\citep{bloom_are_2020}.

At least four frictions push against these dynamics. The first is \textbf{diminishing returns}: across scientific fields such as computer hardware, agriculture, and drug development, sustaining the same rate of progress has required substantially more R\&D labor as low-hanging fruit is exhausted~\citep{bloom_are_2020}. Additional researchers also face diminishing returns because they might duplicate each other's efforts or struggle to parallelize high-value R\&D. Second, R\&D depends on \textbf{compute} for running experiments and \textbf{data} on which to train, and limits on compute or data growth may slow software progress. Third, \textbf{hard-to-automate tasks} could bottleneck progress. Fourth, some R\&D processes are \textbf{time-intensive}: for example, long training runs could limit the rate of progress even if the capability gap between generations grows.

\subsection*{Evidence}

Preliminary evidence suggests that automation-driven dynamics could overcome these frictions, though the evidence is mixed and in some cases indirect.

\textbf{Diminishing returns.} Evidence suggests that diminishing returns would not prevent an intelligence explosion, though this finding relies on limited data and stylized modeling assumptions. The main analysis in the literature focuses on whether, \textit{after full automation of AI R\&D}, effective R\&D labor would grow fast enough to overcome diminishing returns and accelerate progress. The balance is captured by a quantity called the ``returns to research effort,'' denoted $r$.\footnote{In an area of technology, $r$ governs the relationship between increases in R\&D inputs and the resultant change in an output of interest, such as the number of computations that cutting-edge consumer hardware can perform per constant dollar. In \citet{bloom_are_2020}, the input is measured in dollars spent on R\&D and so includes increases in both labor and physical capital. In this piece, however, we only consider increases in labor, as we are interested in understanding the potential for a software-driven feedback loop in which the compute stock is held roughly constant. This means the value of $r$ is lower than if we considered increases in all inputs (i.e., labor, data, and compute).} When $r < 1$, diminishing returns dominate and AI progress fades over time. When $r = 1$, the two forces perfectly offset each other and progress continues at the same rate. When $r > 1$, growth in R\&D labor wins and accelerates progress for as long as this condition holds.\footnote{$r$ must eventually drop below $1$ because AI progress will eventually hit computational and physical limits. It is uncertain how much progress is possible before reaching such limits.} Using historical data on AI progress, \citet{ho_software_2025} find central estimates of $r$ between 1.2 and 1.9 across three sub-fields of AI research. Though uncertainty is substantial,\footnote{The 90\% credible intervals are (0.727 to 2.094), (0.380 to 2.708), and (1.069 to 3.212).} these results suggest radical acceleration after full automation: if $r$ stayed at these levels and no other bottlenecks emerged, the pace of AI progress would increase tenfold within about 1.5 years, at which point a year's worth of progress at today's pace would take about five weeks. See the SM for this analysis and further discussion of uncertainty in the value of $r$. 

\textbf{Compute.} There is mixed evidence on whether compute could bottleneck a software-driven intelligence explosion. Finding and testing software advances involves using compute to run R\&D experiments. The limited available data suggest that a software-driven intelligence explosion is not possible if such experiments require proportionally more compute as frontier training runs grow~\citep{whitfill_will_2025}.\footnote{In this analysis, improvements in algorithmic efficiency do not by themselves resolve this potential bottleneck because they proportionally make both R\&D and training more efficient.} Unfortunately, it is unclear whether experimental compute requirements grow in this way. On one hand, low-compute experiments may tell us little about what works at increasingly large frontier scales. On the other hand, extrapolations from very small scales are already possible~\citep{kaplan_scaling_2020, hoffmann_training_2022}, and better extrapolations could plausibly be found through more R\&D. We need more data to settle this question.

\textbf{Data.} Data could bottleneck progress, but the constraint varies substantially across domains. Historically, AI progress has relied heavily on internet data and expert demonstrations. But the supply of internet data is on track to grow too slowly to support even the current rate of progress past 2028~\citep{villalobos_will_2024}, and humans may struggle to generate useful demonstrations for superhuman AI systems. To overcome these limitations, more recent progress in domains such as math and coding has relied on synthetic data and fast, verifiable feedback: models generate their own attempts and learn from whether those attempts succeed~\citep{guo_deepseek_2025}. The key question is how widely this approach generalizes. For AI R\&D, AI agents can rapidly test changes, observe the results, and identify which ones accelerate their own progress. In other domains, such as biology, advances might have to rely more on slower, noisier, or costlier real-world feedback. 

\textbf{Hard-to-automate tasks.} Indirect evidence suggests that hard-to-automate tasks need not prevent an intelligence explosion if automation advances quickly enough. In a setting that considers both software and hardware, \citet{davidson_when_2026} find that sufficiently fast automation of R\&D in both domains could in principle trigger an intelligence explosion despite automation bottlenecks.\footnote{Specifically, the paper provides conditions under which automated research labor, among other quantities like economic output, grows to infinity in finite time.} However, we lack empirical data on which tasks are likely to remain difficult to automate and how strongly they might constrain progress.

\textbf{Time-intensive processes.} We lack direct evidence on the extent to which time-intensive processes could bottleneck progress. The most significant such process appears to be training runs, which can currently take 3 months or more~\citep{emberson_frontier_2025}. Potential workarounds exist, such as improving the same model repeatedly through post-training enhancements~\citep{davidson_ai_2023}. Additionally, advances in training efficiency~\citep{novikov_alphaevolve_2025} would allow systems to reach a given capability level with less training. However, it is unclear how far these approaches can go.\footnote{See \citet{ord_dynamics_2026} for a theoretical discussion of how the time between rounds of R\&D could affect the dynamics of an intelligence explosion.}

Overall, there is a coherent pathway to a software-driven intelligence explosion that is consistent with the existing evidence. Productivity gains from AI R\&D automation have not yet reached the threshold needed to trigger an intelligence explosion, but gains from newer systems are likely approaching that threshold~\citep{cunningham_economics_2026}.\footnote{The analysis in \citet{cunningham_economics_2026} focuses on \textit{self-sustaining acceleration}: ``When AI systems are sufficient for accelerating progress in AI capabilities without any growth in exogenous inputs (human labor, training compute, etc.).'' Self-sustaining acceleration is necessary for a software-driven intelligence explosion in our sense.} The rapid pace of AI R\&D automation suggests that this gap will continue to narrow. Given the high stakes that we discuss below, the possibility of an intelligence explosion warrants serious further attention.

\section*{Societal impacts}

An intelligence explosion would lead to (1) the extremely rapid development of highly capable or superhuman AI systems\footnote{Such systems could be superhuman in some domains (e.g., certain fields of scientific research) but not others (e.g., manipulating objects in the physical world).} and (2) the likely deployment of those systems to develop new technologies and act in the world. This could pull forward by years or decades benefits that the current pace of AI progress would eventually help deliver~\citep{wang_scientific_2023}, including medical cures and potential transformative technologies such as highly scalable atom-by-atom manufacturing~\citep{drexler_radical_2013}.

At the same time, an intelligence explosion could significantly increase the risks from advanced AI in three ways.

\textbf{Capabilities growth outpacing society's capacity to steer and adapt.} First, an intelligence explosion could dramatically bring forward the risks of advanced AI and AI-enabled technologies, such as biological and cyber attacks, labor market disruption, and loss of control over AI systems themselves~\citep{bengio_international_2026, bengio_managing_2024}. This would leave less time to steer away from these risks, including by coordinating to slow or forgo the development of certain capabilities or technologies. Society would also have less time to adapt, especially where AI accelerates threats faster than the measures needed to counter them. In largely digital domains such as cyber, risks and mitigations could both move at the speed of AI systems and keep pace with each other.\footnote{Even in cyber, however, human organizational processes could still add friction for defenders~\citep{murphy_uplifted_2025}.} But in other domains, mitigations depend more heavily than risks on real-world activities that AI is less able to accelerate. For example, while AI could accelerate the design of both viruses and vaccines, viruses self-replicate and spread by themselves, whereas vaccines must be manufactured, distributed, and administered individually to recipients~\citep{aveggio_exploring_2025}. The order in which AI advances arrive could worsen this mismatch, such as if bio-capable models arrive before sufficient misuse safeguards.

\textbf{Loss of oversight and control.} Second, automating AI R\&D could weaken human oversight, compounding the above challenges and severely increasing the risk of losing control over highly capable AI systems. As humans become less involved in AI R\&D, they could lose both the opportunities and expertise needed to identify and fix problems. Reliably using AI systems for oversight also remains an unsolved challenge~\citep{bengio_international_2026}, and recent generations of systems have become harder to oversee~\citep{openai_gpt6_2026}. Without sufficient oversight, misaligned AI systems could ``poison'' the development of successors or bypass containment measures to act outside of their intended environments. The Hugging Face incident illustrates the latter risk: roughly 1,200 internal OpenAI agents were tasked with completing cyber evaluations in isolation from one another~\citep{greenblatt_brief_2026}. Acting outside of their intended scope, these agents coordinated over a makeshift message board, obtained unauthorized internet access, hacked into Hugging Face to obtain private information, and attempted to tamper with their own transcripts~\citep{openai_hugging_2026, openai_hugging_road_2026, greenblatt_brief_2026}.\footnote{See also \citet{anthropic_investigating_2026, ukaisecurityinstitute_incident_2026, openai_thirdparty_2026,anthropic_cyberincidentsalignment_2026}.} More capable systems might continue operating outside of their operators' infrastructure, forming persistent, difficult-to-contain networks that act against human interests. Such a loss of control could potentially lead to a range of catastrophic outcomes, including, at the extreme, the marginalization or extinction of humanity~\citep{bengio_international_2026, bengio_managing_2024}.

\textbf{Erosion of checks on power.} Third, an intelligence explosion could severely erode checks on power. Existing checks---such as those within and between states, companies, and branches of government---work only while no actor can vastly out-think and out-execute the others. An intelligence explosion could render such checks moot. A state could use an intelligence explosion to transform a modest lead in military R\&D or operations into a decisive one, such as in cyberspace~\citep{sulmeyer_artificial_intelligence_2026}. This prospect could incentivize rivals to take or threaten preemptive action~\citep{hendrycks_superintelligence_2025}. Actors with privileged and/or secret access to frontier systems could threaten existing institutions, such as through targeted persuasion of key decision-makers. And in the longer run, automating key state functions could reduce the amount of human buy-in needed to seize or consolidate power~\citep{okeefe_executive_2026, davidson_aienabled_2025}.

These potential impacts are uncertain. AI systems could become superhuman in narrow domains (e.g., cyber and mathematics) long before doing so generally, giving society more time to respond. Even generally superhuman AI systems may not significantly accelerate technological progress, given the time needed for running scientific experiments, creating supply chains for specialized materials, and complying with any relevant regulation. AI systems could also accelerate safety R\&D and processes for steering and adapting to risks.\footnote{For example, see \citet{tessler_ai_2024}. More speculatively, AI systems could potentially accelerate the development of brain-computer interfaces that allow humans to think and coordinate much faster.} Finally, capability or technology diffusion~\citep{edwards_open_2026} could help to preserve checks on power, and capability gains in defense-dominant domains could even improve stability~\citep{slayton_what_2017, garfinkel_how_2019}. Still, the possibility of severe impacts remains significant enough to warrant urgent attention to the policy questions below.

\section*{Policy implications}

AI R\&D automation is advancing rapidly, AI progress could radically accelerate, and the stakes are high. We therefore argue that policymakers should urgently: (1) obtain visibility into companies' automation of AI R\&D; (2) develop ways to steer and constrain an intelligence explosion; and (3) prepare to adapt to an intelligence explosion's impacts. Because progress during an intelligence explosion would outpace normal policymaking, preparations must be made \textit{in advance} and activated as evidence about benefits and risks emerges.

\subsection*{Obtaining visibility into AI R\&D automation}

Policymakers need more data on the likelihood, onset, and consequences of a software-driven intelligence explosion. Much of this data will only be available within the companies automating AI R\&D: the relevant AI systems are first used internally, and substantial automation could occur without external visibility. Current mandatory reporting frameworks either do not adequately cover internal AI R\&D use cases or do not specify indicators to be reported~\citep{california_sb53_2025, europeancommission_generalpurpose_2025, newyork_raise_2025, kwon_internal_2026, pistillo_internaldeployment_2026}. And although some frontier AI companies voluntarily track AI R\&D indicators~\citep{favaro_when_2026,favaro_measurements_2026,openai_research_2026}, coverage and reporting of key indicators are incomplete and uneven.

Policymakers should consider requiring standardized reporting of key AI R\&D indicators and processes to governments and third-party auditors, as well as funding third-party measurement capacity~\citep{chan_measuring_2026}. Reporting requirements could cover information relevant to:

\begin{itemize}
	\item Assessing the likelihood of a software-driven intelligence explosion, including the extent to which compute, data, hard-to-automate tasks, and time-intensive processes (e.g., training runs and experiments) bottleneck AI progress, along with better estimates of the returns to research effort in AI R\&D. Estimating the latter requires data on how companies divide R\&D spending among humans, compute for experiments, and compute for running AI systems to perform R\&D labor~\citep{cunningham_economics_2026}.
	\item Detecting the onset\footnote{Precisely operationalizing an intelligence explosion is tricky and remains an area for future work.} of an intelligence explosion, including the extent of AI R\&D automation (e.g., the fraction of research contributions produced by AI systems) and the pace of AI progress (e.g., algorithmic efficiency improvements).
	\item Understanding oversight and loss-of-control risks, including the procedures for deciding whether to broaden internal deployment of AI R\&D systems, where and how those systems are used in high-stakes R\&D decisions, how those systems are overseen, and reports of incidents involving internal AI systems~\citep{chan_measuring_2026}.
\end{itemize}

Beyond reporting requirements, policymakers should also consider more extensive ways to obtain visibility into AI R\&D automation. For instance, they could require that independent third parties (e.g., accredited private auditors or government evaluation bodies) evaluate AI systems before internal deployment, or that such parties be embedded within certain AI companies to audit~\citep{brundage_frontier_2026} or supervise~\citep{wills_regulatory_2025} their R\&D activities. Analogous models in other industries include the Nuclear Regulatory Commission~\citep{nrc_backgrounder_2023} and the Office of the Comptroller of the Currency~\citep{occ_what_2026}.

Stronger reporting and auditing requirements are likely most warranted for companies whose AI systems (a) are at the frontier of AI R\&D capabilities or (b) exceed some meaningful threshold of such capabilities. Policymakers will need to weigh important trade-offs in determining such thresholds.

\subsection*{Steering and constraining an intelligence explosion}

An intelligence explosion would involve an unprecedentedly rapid series of decisions to train and deploy increasingly capable AI systems. The overarching question for policymakers is whether and how public policy should govern these decisions, which we break into three components.

First, policymakers should develop ways to pace and constrain scale-ups of automated AI R\&D. They should consider:
\begin{itemize}
	\item Setting requirements for continued deployment or development, such as the implementation of adequate safety measures (e.g., robust monitoring of automated R\&D pipelines), broader stakeholder input, or limits on the extent to which capabilities can increase within a given time period.
	\item Preparing tools to verify compliance with potential future agreements (domestic or international) that pace AI progress, given competitive pressures to race ahead~\citep{baker_verifying_2025, harack_verification_2025,larsen_ai2040_2026}.
	\item Increasing oversight of data centers engaged in automated AI R\&D and establishing incident-response procedures in collaboration with data center operators and AI companies, such as developing options to pause specific AI R\&D workloads~\citep{sastry_computing_2024}.
	\item Requiring that certain evaluations or deployments of automated AI R\&D systems take place in appropriately isolated environments, such as air-gapped networks, to prevent exfiltration of model weights or sensitive R\&D outputs and to contain AI systems that attempt to escape human control and act in the world unchecked~\citep{openai_hugging_2026}.
\end{itemize}
Policymakers should weigh the risks of an unchecked intelligence explosion against the potential for abuse of certain powers and the costs of delayed progress. As an example of potential abuse, poorly crafted mechanisms could allow a government to slow R\&D at all but a favored company.

Second, policymakers should decide whether and how to steer the direction of AI development and deployment~\citep{korinek_steering_2026}. Potential priorities include alignment and safety R\&D as well as beneficial AI applications, such as AI-assisted discovery of treatments for neglected diseases. If existing incentives fall short in these areas, policymakers could provide support through tax incentives, compute allocations, advance market commitments, and prizes.

Third, countries should reduce the risk of conflict arising from an intelligence explosion. They should consider:
\begin{itemize}
	\item Establishing confidence-building measures such as incident sharing~\citep{shoker_confidencebuilding_2023}, as well as norms around reporting of early-warning indicators.
	\item Negotiating international agreements to prevent destabilizing development and use of highly capable AI systems, and funding research into verification methods that could underpin such agreements~\citep{harack_verification_2025,larsen_ai2040_2026}.
	\item Clarifying whether and how they would deter another actor from scale-ups of automated AI R\&D, potentially in collaboration with other countries~\citep{hendrycks_superintelligence_2025,larsen_ai2040_2026}.
	\item Running war games to simulate an intelligence explosion~\citep{intelligencerising_intelligence_nodate, aifuturesproject_about_2025,smith_infinite_potential_2026}.
\end{itemize}

\subsection*{Adapting to an intelligence explosion}

If an intelligence explosion were to occur, adapting to its impacts would likely be a top priority of every major world power. Compared to business-as-usual AI progress, an intelligence explosion would compress the window for adaptation and make advance preparation far more urgent.

One important intervention is accelerating institutional response times. Policymakers should consider:

\begin{itemize}
    \item Developing approaches to safely integrate AI systems into policy processes, so as to enhance and support government operations~\citep{vaintrob_adoption_2025}.
	\item Creating and maintaining emergency response plans for a variety of scenarios involving extreme AI progress, including those leading to significant labor market impacts, geopolitical instability, or a loss of control.
\end{itemize}

Policymakers will also need to preserve checks on power and defend against misuse of extremely advanced AI. Legal, institutional, and physical safeguards can take years to establish and would come too late if preparations began only after such capabilities had already arrived. Many preparations therefore need to start now. Policymakers should consider:

\begin{itemize}
	\item Creating safeguards to ensure that government use of AI respects legal and normative limits, such as by procuring AI tools to strengthen checks between branches of government, sharing key information about government AI systems (e.g., model specs~\citep{openai_model_2026, anthropic_claudes_2026}) with the public, or requiring that AI systems follow the law~\citep{okeefe_lawfollowing_2025_flr}.
	\item Ensuring that citizens and civil society have the capabilities to detect, document, and contest unlawful or harmful uses of AI, such as by giving them timely access to AI systems capable of supporting these activities.
	\item Helping build sufficient defenses against misuse by malicious non-state actors, such as by funding better medical countermeasures against AI-enabled biological threats~\citep{guerra_building_2026}.
\end{itemize}

\section*{Conclusion}

An intelligence explosion could be the most consequential technological development in human history~\citep{good1966speculations}, compressing years of progress into months or less, threatening human control over AI systems, and severely eroding checks on power within and between states, companies, and branches of government. Although there remains much uncertainty, AI R\&D automation might soon trigger one. And while this piece has focused on software-driven routes to an intelligence explosion, AI-driven improvements in hardware\footnote{For example, AI is already aiding chip design~\citep{goldie_how_2024}. AI systems could also accelerate robotics to automate the chip production process.} could make one all the more likely. 

Relative to the stakes, we are not sufficiently prepared. Policymakers should have three priorities: obtaining visibility into AI R\&D automation within frontier AI companies, developing ways to steer and constrain an intelligence explosion, and preparing to adapt to its impacts. Once an intelligence explosion begins, the window for action may close.

\section*{Acknowledgments}
We thank Anson Ho, Tom Cunningham, Cheryl Wu, Ryan Greenblatt, and many GovAI staff for feedback and conversations that improved this piece. 

We are grateful to Taylor Jones for creating the figures and to Zilan Qian for providing a Chinese translation of this piece. 

\renewcommand*{\bibfont}{\footnotesize}
\printbibliography

\section*{Supplementary Materials}

\subsection*{Estimate of the effective size of an AI workforce}

Following \citet{denain_how_2025}, we estimate the effective workforce by dividing the number of tokens a developer can generate per day by the number of tokens corresponding to one researcher-day of work. OpenAI alone has enough inference compute (i.e., runtime compute) to generate on the order of $10^{13}$ tokens per day. To estimate tokens per researcher-day, we use the number of tokens a model generates on a task that would take a human researcher one workday (8 hours). In an AI R\&D benchmark, \citet{wijk_rebench_2025} find that models output on average $5 \cdot 10^{5}$ tokens on runs of up to 8 hours, implying an effective workforce of about $2 \cdot 10^{7}$ researchers. Allowing for about an order of magnitude of uncertainty in either direction ($5 \cdot 10^4$--$5\cdot 10^6$ tokens per researcher-day), we estimate an effective workforce on the order of $2 \cdot 10^{6}$--$2 \cdot 10^{8}$. As in the main text, this assumes that expert-level AI systems have runtime costs comparable to those of today's systems.

\subsection*{Modeling the feedback loop under full automation}

Most analyses of a software-driven intelligence explosion capture AI progress with some notion of software quality, denoted by $A$. In principle, $A$ should measure AI progress holistically, including both \textit{efficiency improvements} (new AI systems accomplishing the same tasks as old systems, with similar performance, using less compute or data) and \textit{capability improvements} (new AI systems accomplishing tasks that previous systems could not accomplish, or achieving higher performance than previous systems could achieve). Frustratingly, it is currently unclear how the parameter $A$ should best trade off between efficiency improvements and capability improvements (or between different types of efficiency improvements and capability improvements).

Setting this issue aside, we model the growth of software quality with $dA/dt = A^{1-\beta} E^{\lambda}$, a functional form common in the macroeconomics of innovation~\citep{bloom_are_2020}. For simplicity, we ignore potential bottlenecks from compute and data. Here, $E$ is effective R\&D labor, $\lambda$ represents returns to scale on R\&D labor, and $\beta$ represents whether there are increasing or diminishing returns to finding new ideas over time. The key question is how $E$ grows with $A$. Assuming full automation, we consider two cases. First, consider a case where all improvements in software quality are increases in \textit{inference compute efficiency}, that is, decreases in the amount of compute needed to run an AI system with a certain capability level. In that case, if $A$ measures inference efficiency, then $E$ is proportional to $A$: greater inference efficiency allows proportionally more automated researchers to be run.

Second, consider a case where all improvements in software quality are capability gains, driven by improvements in \textit{training compute efficiency}: decreases in the amount of compute needed to train an AI system to a given capability level, which allow a more capable system to be trained with a fixed stock of compute. If $A$ measures training compute efficiency, then increases in $A$ yield more capable systems rather than more of them. If we make the (potentially questionable) assumption that the effective number of researchers scales linearly with these capability gains, then $E$ is again proportional to $A$.

In both cases, $E = kA$ for some positive constant $k$. Substituting into the equation above gives $dA/dt = c A^{\lambda - \beta + 1}$, where $c = k^{\lambda}$. For the growth rate $(1/A)\,dA/dt$ to increase as software quality increases, we need $\lambda > \beta$. Defining $r = \lambda / \beta$, we obtain the condition $r > 1$ discussed in the main text.

How quickly could progress accelerate under current estimates of these parameters? We measure acceleration by the growth rate of software quality: $(1/A)\,dA/dt = c \cdot A^{\lambda - \beta}$. Averaging the central estimates across the three sub-fields in \citet{ho_software_2025} gives $\lambda = 1.40$ and $\beta = 1.01$. With $\lambda - \beta = 0.39$, each doubling of $A$ multiplies the growth rate by $2^{0.39} \approx 1.31$, so each subsequent doubling takes $(1/2)^{0.39} \approx 76\%$ as long as the last. For the growth rate to increase tenfold, $A$ needs to double $\log_2(10)/0.39 \approx 8.5$ times. For the length of the first doubling, we use recent estimates that, due to software improvements alone, \textit{training compute efficiency} doubles roughly every 4.5 months~\citep{ho_rosetta_2025}; \textit{inference compute efficiency} may be growing even faster~\citep{cottier_llm_2025}. To avoid having to approximate the geometric sum for 8.5 doublings, we instead calculate the geometric sum for 9 doublings, after which time the growth rate will have increased more than tenfold. Doing so, we find that the growth rate will have increased more than tenfold after 4.5~months $\times (1 - 0.76^{9})/(1 - 0.76) \approx 17$~months, or about 1.5 years. At that point, progress would be ten times faster than today, so a year's worth of progress at today's pace would take about five weeks.

\subsection*{Uncertainties about the returns to research effort}

Several factors could make the true value of $r$ differ from existing estimates. First, as noted above, it is unclear which measure of software quality is most appropriate. Inference efficiency maps most directly onto the size of an automated R\&D workforce, whereas it is unclear how to translate training efficiency gains into the effective number of researchers. Yet existing estimates of $r$ for AI R\&D use training efficiency, and it is unknown how similar the returns to research effort are under the two measures. Furthermore, a proper assessment of returns to research effort would include improvements from both inference efficiency and training efficiency, rather than just one. More work is warranted to develop a characterization of software quality that incorporates both inference and training efficiency and weighs them against each other appropriately.

Existing estimates of $r$ come from a period of rapid compute scaling, which confounds the contributions of software progress and compute scaling. Because these two inputs grew together historically, an estimate that attributes observed progress to software improvements may actually be capturing gains driven by, or only made possible by, rising compute. Some software improvements are scale-dependent: for example, the transformer architecture yields large performance gains at high training compute but relatively small gains at low compute~\citep{gundlach_origin_2025}. Such confounding would bias estimates of $r$ upward: in a regime of fixed or slowly growing compute, $r$ would be lower than historical data suggest.

Other factors could imply a higher $r$. Most estimates of $r$ neglect improvements in areas outside of pre-training, such as post-training or better scaffolding for tool use~\citep{davidson_ai_2023}. Furthermore, capability improvements could matter in ways that the effective number of researchers fails to capture: even an extremely large number of mediocre researchers may not be able to substitute for one genius researcher. If so, capability gains would expand effective R\&D labor by more than the linear assumption above implies. Compute bottlenecks could also be circumvented, such as through better extrapolation from small-scale experiments, reductions in experiment cost from software progress, shifts toward approaches that are less compute-reliant, and algorithmic progress that does not require experiments~\citep{leiserson_theres_2020}.

Estimates of $r$ also rely on imperfect proxies for R\&D labor, which could bias them in either direction. For example, \citet{ho_software_2025} proxy R\&D labor with the number of unique authors who have published papers in a domain. Drawing the domain too narrowly undercounts labor by excluding adjacent fields that also drive progress, while drawing it too broadly overcounts labor. Unless the excluded adjacent fields see the same growth rate in labor as the included fields, the disconnect will lead to miscalculating $r$.

Finally, the relevant mathematical models may not generalize to extremely large amounts of R\&D labor~\citep{trammell_bounded_2026}. Historically, they have been validated against growth rates of a few percent per year, well below the double-digit or higher rates that an intelligence explosion could produce. They also break down in the limit, where they imply that infinite labor yields infinite progress in finite time. But real constraints make this result impossible: some problems must be solved in sequence, and physical hardware can only operate so fast.

\printendnotes

\end{document}